\documentclass[%
reprint,
amsmath,amssymb,
aip,
]{revtex4-1}

\usepackage{graphicx}% Include figure files
\usepackage{dcolumn}% Align table columns on decimal point
\usepackage{bm}% bold math
\usepackage{animate}

\usepackage{color}

\usepackage{mathtools}
\usepackage{makecell,tabularx}
\setcellgapes{3pt}

\usepackage{soul}

\begin{document}

%\preprint{APS/123-QED}

\title{Supervised Chaotic Source Separation by a Tank of Water}% Force line breaks with \\

\author{Zhixin Lu}
\author{Jason Z. Kim}
\author{Danielle S. Bassett}
\email{dsb@seas.upenn.edu}
\altaffiliation[Also at ]{%
Also Departments of Physics \& Astronomy, Electrical \& Systems Engineering, Neurology, and Psychiatry at the University of Pennsylvania, Philadelphia, PA 19104; and the Santa Fe Institute, Santa Fe NM 87501
}%

\affiliation{Department of Bioengineering, School of Engineering \& Applied Sciences, University of Pennsylvania, Philadelphia, PA, 19104}%Lines break automatically or can be forced with \\

%\date{\today}

\begin{abstract}
Whether listening to overlapping conversations in a crowded room or recording the simultaneous electrical activity of millions of neurons, the natural world abounds with sparse measurements of complex overlapping signals that arise from dynamical processes. While tools that separate mixed signals into linear sources have proven necessary and useful, the underlying equational forms of most natural signals are unknown and nonlinear. Hence, there is a need for a framework that is general enough to extract sources without knowledge of their generating equations, and flexible enough to accommodate nonlinear, even chaotic, sources. Here we provide such a framework, where the sources are chaotic trajectories from independently evolving dynamical systems. We consider the mixture signal as the sum of two chaotic trajectories, and propose a supervised learning scheme that extracts the chaotic trajectories from their mixture. Specifically, we recruit a complex dynamical system as an intermediate processor that is constantly driven by the mixture. We then obtain the separated chaotic trajectories based on this intermediate system by training the proper output functions. To demonstrate the generalizability of this framework \emph{in silico}, we employ a tank of water as the intermediate system, and show its success in separating two-part mixtures of various chaotic trajectories. Finally, we relate the underlying mechanism of this method to the state-observer problem. This relation provides a quantitative theory that explains the performance of our method, such as why separation is difficult when two source signals are trajectories from the same chaotic system.
\end{abstract}

\maketitle

%\tableofcontents
\begin{quotation}
Experimentally measured signals from the natural world are often mixtures of dynamical signals that are generated by independently evolving sources systems. The chaotic source separation (CSS) problem is to infer the separated signals, each of which is generated by a chaotic system, given the measured signal comprised of their mixture. In this paper, we show that the chaotic source separation problem can be considered as a nonlinear state-observer problem, and we propose a dynamical framework with notable generalizability to solve the CSS problem without knowing the governing dynamical equations of the source systems. As a demonstration, we recruit a tank of water as an intermediate processor that is driven by the mixture signal. We show that through supervised training, the separated signal can be obtained from the states of the tank of water. By relating chaotic source separation to the nonlinear state-observer problem, we also explain why separation is difficult when the two source signals are trajectories from the same chaotic system.  More broadly, our study provides a foundation for the principled examination of source separation in nonlinear dynamical data.
\end{quotation}

\section{\label{sec:intro}Introduction}

Blind source separation (BSS) is the separation of source signals from a mixed signal with little or no information regarding the source signals or the mixing process. A classical example is the cocktail party problem, where a listener follows any one of many simultaneously occurring conversations at a cocktail party. BSS also has many notable applications in digital signal processing, such as removing artifacts from electroencephalography (EEG) and magnetoencephalography (MEG) recordings \cite{vigario2000independent,jung2000removing,zhou2004removal,fitzgibbon2007removal,vazquez2012blind}. When the mixed signal has a lower dimension than the total dimension of the sources, the BSS is called under-determined. 

Many methods have been proposed to solve BSS in various scenarios. For example, by assuming various types of statistical independencies or mixing properties of the source signals, unsupervised classical methods such as principal component analysis (PCA)\cite{karhunen1995nonlinear,stetter2000principal}, independent component analysis (ICA) \cite{comon2010handbook,pham1996blind}, and non-negative matrix factorization (NMF)\cite{cichocki2006new,wang2005musical} have been proposed. While these methods have dramatically enhanced our ability to parse data from linear and static statistical distributions, it has been shown that adaptations of the above methods \cite{phon2013handling,he2018single,sawada2019review} as well as many other supervised learning methods, such as the Wiener filter \cite{sharma2015blind}, support vector machines \cite{han2012classification}, deep learning networks\cite{mogami2018independent,wang2018supervised}, and recurrent neural networks \cite{lu2018parsimonious,krishnagopal2019separation}, outperform classical methods when the signals are generated from complex dynamical sources.

In this paper, we focus on a particular type of separation problem: chaotic source separation (CSS). Specifically, we consider the $d$-dimensional mixed signal to be a superposition of two $d$-dimensional trajectories, each of which is generated by an autonomous $d$-dimensional chaotic system. This problem is of particular relevance in high-dimensional biological signals such as those from chaotic neural systems, as experimental measurements involve a mixture of electrical activity, correlated artifacts, and hemodynamic response \cite{jung2000removing}. Hence, it is of interest to study how one can extract a chaotic trajectory of interest from the mixed signal. 

Here, we propose to solve this problem with an intermediate dynamical system that is trained by a supervised learning method. Although the dimension of the mixed signal $d$ is only half of the total dimension $2d$, the problem can still be solved by a supervised learning framework, where the exact separated trajectories are known during a training period. As a significant extension from previous works \cite{andreyev2003separation,tsimring1996multiplexing} that require knowledge of the governing equation of the source chaotic systems, we build on a prior demonstration from the present authors that a recurrent neural network (RNN) can solve the CSS problem in the absence of these equations \cite{lu2018parsimonious} (a more recent work by Krishnagopal \emph{et al.} also demonstrated that a reservoir computer can solve the CSS problem \cite{krishnagopal2019separation}). In this paper, we extend the demonstration by enacting this separation through a dynamically simple intermediary system that is a simulated tank of water, and provide a quantitative theory explaining why and how such chaotic source separation is solvable. Our theory accurately predicts that separation is harder when the two source signals are generated by the same chaotic system, and provides a foundation for the principled study of source separation in nonlinear dynamical data.

\section{\label{sec:model}Supervised learning model for chaotic source separation}

\subsection{\label{sec:scheme}General scheme}
\begin{figure}
	\includegraphics[width=0.48\textwidth]{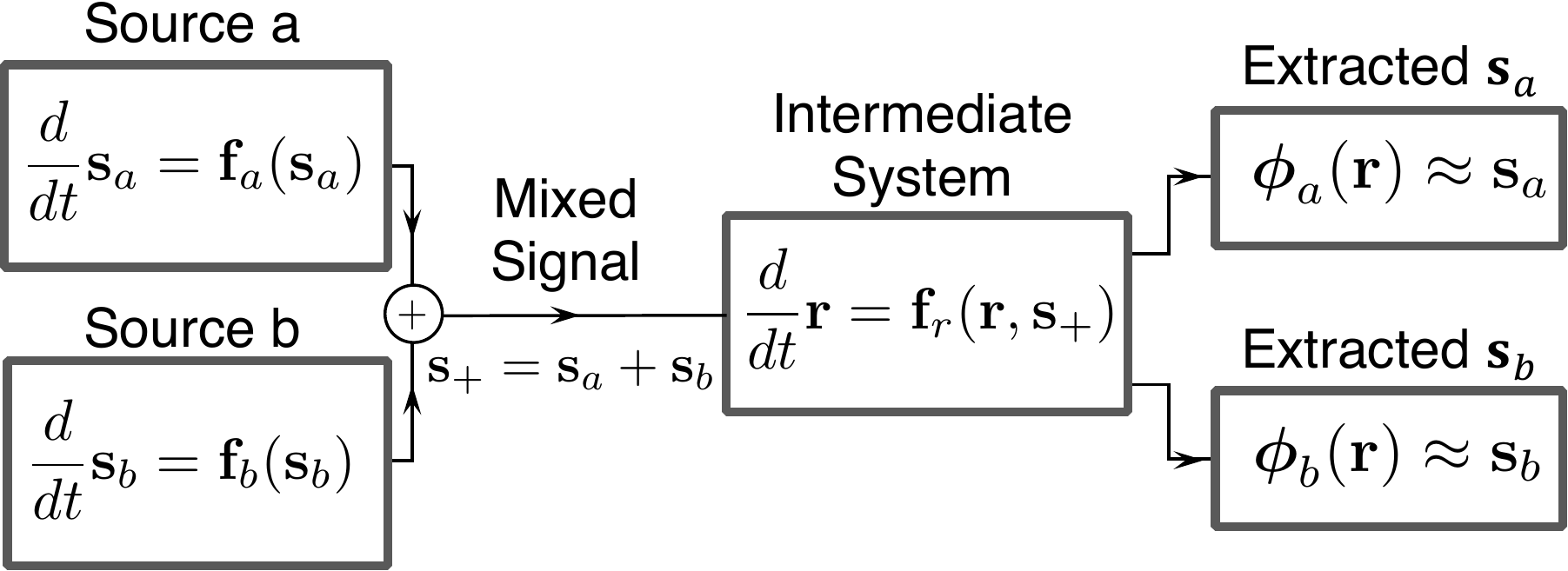}
	\caption{Schematic plot of the source separation model. Two signals, $\mathbf{s}_a(t)$ and $\mathbf{s}_b(t)$ are generated independently by the two sources, which are two chaotic dynamical systems. The two signals are mixed into $\mathbf{s}_+(t)=\mathbf{s}_a(t)+\mathbf{s}_b(t)$. The intermediate system evolves with $\mathbf{s}_+$ as its driving signal. Supervised learning is applied to find proper readout functions ${\bm \phi}_a$ and ${\bm \phi}_b$ that extract the separated signals $\mathbf{s}_a$ and $\mathbf{s}_b$ from their mixture $\mathbf{s}_+$.}
	\label{fig:schm}
\end{figure}
We begin with a simple description of a general scheme of our supervised CSS (Fig.~\ref{fig:schm}). We consider extracting trajectories of two autonomously evolving chaotic systems, $\mathbf{s}_a(t)$ and $\mathbf{s}_b(t)$, from their mixture, $\mathbf{s}_+(t)=\mathbf{s}_a(t)+\mathbf{s}_b(t)\in\mathbb{R}^d$, where
\begin{subequations}
\begin{align}
\dot{\mathbf{s}}_a(t)&=\mathbf{f}_{a}(\mathbf{s}_a), \\
\dot{\mathbf{s}}_b(t)&=\mathbf{f}_{b}(\mathbf{s}_b).
\end{align}
\label{eqn:source_systems}%
\end{subequations}
CSS is similar to an underdetermined BSS problem in the sense that the dimension of the mixture $d$ is less than the total dimension of the sources, i.e., $2d$. As a result, there exist mixed states $\mathbf{s}_+$ that correspond to multiple distinct pairs of sources. Thus, without utilizing the temporal structure, one cannot find a function that maps the simultaneous state $\mathbf{s}_+(t)$ to the separated states $\mathbf{s}_a(t)$ and $\mathbf{s}_b(t)$.

The essential idea of our method is to implement a high-dimensional dynamical system as an intermediate system
\begin{equation}
\frac{d}{dt}\mathbf{r}=\mathbf{f}_r(\mathbf{r},\mathbf{s}_+),
\label{eqn:intermediate}
\end{equation}
that is continuously driven by $\mathbf{s}_+(t)$. Since the state of the intermediate system, $\mathbf{r}$, incorporates both the immediate value and the history of the mixed signal $\mathbf{s}_+(t)$, one may obtain the full states of the two sources by training the output functions ${\bm \phi}_a(\cdot)$ and ${\bm \phi}_b(\cdot)$, as shown in Fig.~\ref{fig:schm}. We assume that the governing equations of the source systems are unknown. However, different from the BSS problem, we do assume that the separated trajectories $\mathbf{s}_a(t)$ and $\mathbf{s}_b(t)$ are known for a finite time window. During this time window, we match the recorded state of the intermediate system $\mathbf{r}(t)$ with the two separated signals $\mathbf{s}_a(t)$ and $\mathbf{s}_b(t)$, and we look for two functions, ${\bm \phi}_a(\cdot)$ and ${\bm \phi}_b(\cdot)$, which can estimate the separated signals based on the state of the intermediate system, i.e., ${\bm \phi}_a(\mathbf{r}(t))\approx \mathbf{s}_a(t)$ and ${\bm \phi}_b(\mathbf{r}(t))\approx \mathbf{s}_b(t)$ where $t$ is in the time window.

\subsection{\label{sec:water}Intermediate system instantiated by a tank of water}
\begin{figure}
	\center
	\includegraphics[width=0.48\textwidth]{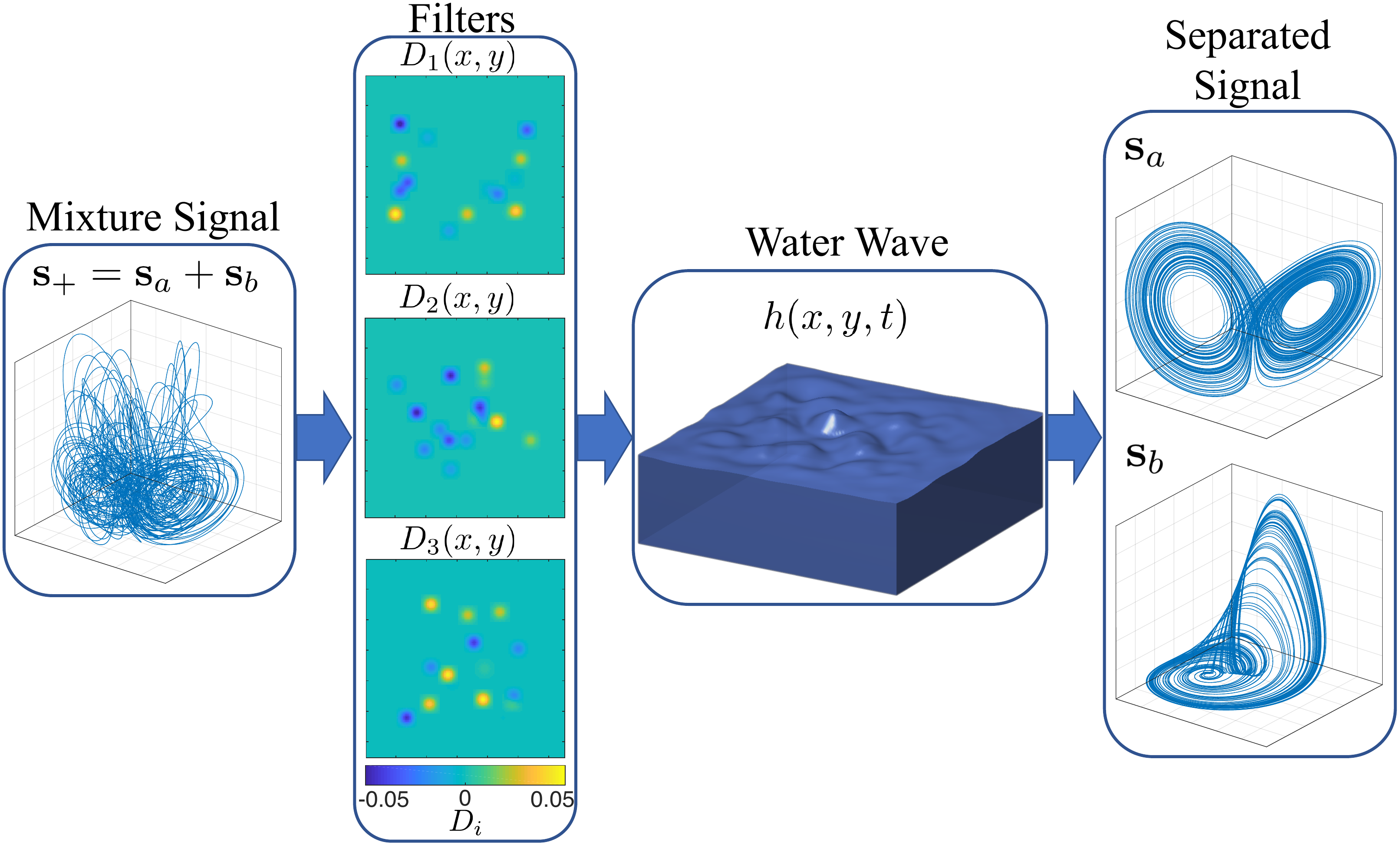}
	\caption{A schematic demonstration of chaotic source separation performed by a tank of water. The mixture signal is the sum of two chaotic trajectories, one from the Lorenz system and the other from the R\"ossler system. This $3$-dimensional mixture signal is then propagated onto the surface of the water wave through three randomly generated filters: $D_i(x,y)$ with $i=1,2,3$. The output function is then trained to estimate the separated signals $\mathbf{s}_a$ and $\mathbf{s}_b$ by measuring the shape of the wave surface.}
	\label{fig:water_animate}
\end{figure}
\begin{figure}
	\centering
	\includegraphics[width=0.45\textwidth]{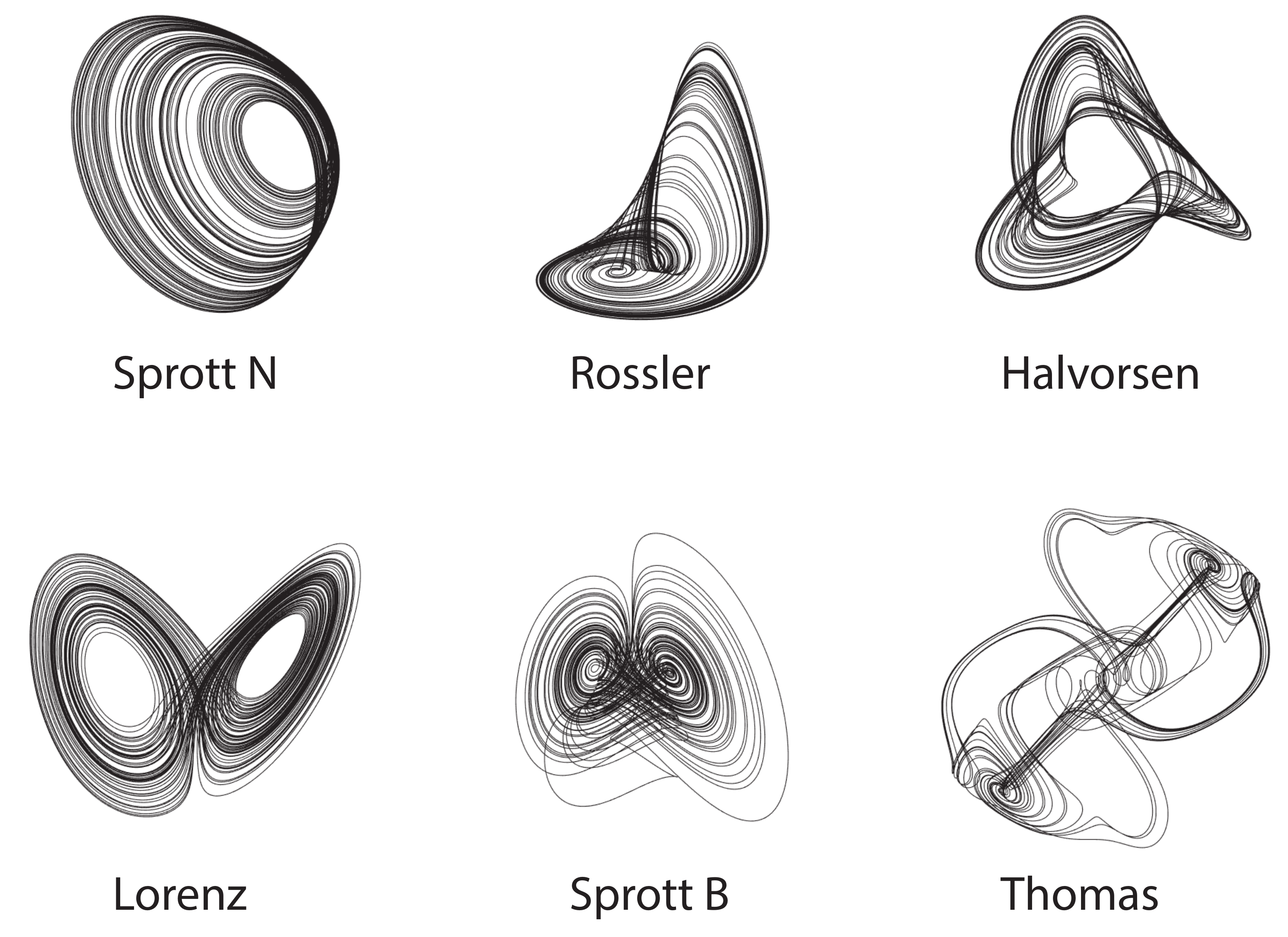}
	\caption{\textbf{Example trajectories of the chaotic attractors studied in this work.} Here we show six exemplary trajectories with duration $5000$ time units on the $6$ chaotic attractors, whose equations are shown in Tab.~\ref{tab:attractors}. We integrate these chaotic attractors using a $4$-th order Runge-Kutta method with a time step of $0.0001$ units.}
	\label{fig:attractor}
\end{figure}

To demonstrate the generalizability of this scheme beyond the RNN used in prior work \cite{lu2018parsimonious}, we instantiate the intermediate system \emph{in silico} as a tank of water (Fig.~\ref{fig:water_animate}). We test the performance of this intermediate system on the CSS problem given the mixed signals of different pairwise sums from $6$ distinct chaotic systems. We show their governing equations in Tab.~\ref{tab:attractors} and their attractors in Fig.~\ref{fig:attractor}. We notice that trajectories of different chaotic systems have different ranges; to simplify the simulation and ensure an accurate quantification of the error, we deliberately preprocess the chaotic trajectories such that all variables have zero mean and unit variance along the time axis.

We construct the intermediate system as a square tank of water that is constantly perturbed by the mixed signal $\mathbf{s}_+(t)$. The perturbed water evolves following the nonlinear partial differential equations
\begin{subequations}
\begin{align}
%\begin{split}
\frac{\partial h}{\partial t} + \frac{\partial uh}{\partial x} + \frac{\partial vh}{\partial y} &= p,
\\
\frac{\partial (uh)}{\partial t} + \frac{\partial (u^2h +\frac{1}{2}gh^2)}{\partial x} + \frac{\partial (uvh)}{\partial y}  + buh & = 0,
\\
\frac{\partial (vh)}{\partial t} + \frac{\partial (uvh)}{\partial x} + \frac{\partial (v^2h +\frac{1}{2}gh^2)}{\partial y} + bvh & = 0,
%\end{split}
\end{align}
\label{eqn:shallow_water}%
\end{subequations}
where $h(x,y,t)$ is the height of the wave surface, $u(x,y,t)$ and $v(x,y,t)$ are the zonal and meridional speeds, respectively, $g=9.8$ is the gravity constant, and $b>0$ is the viscous drag coefficient. The tank has a flat $1$-by-$1$ bottom and four vertical hard walls with reflective boundary condition. When the perturbing term $p=0$, Eqs.~(\ref{eqn:shallow_water}a)-(\ref{eqn:shallow_water}c) become the traditional shallow water equations with the presence of a viscous dragging force \cite{vreugdenhil2013numerical}.

Although other forms of perturbation exist, for example a time-varying bottom which requires the modification of all three equations, for the simplicity of the demonstration, we drive the water by artificially defining the perturbing term
\begin{equation}
p(x,y,t)=\sum_{i=1}^d D_i(x,y)[\mathbf{s}_+(t)]_i,
\label{eqn:filter}
\end{equation}
on the right hand side of Eq.~(\ref{eqn:shallow_water}a) only. The perturbation term $p(x,y,t)$ can be considered to reflect the speed of with which one vertically and inhomogeneously adds or removes water from right above the wave surface. Each component of the $d$-dimensional mixed signal $[\mathbf{s}_+(t)]_i$ is propagated onto the wave surface through randomly constructed input filters, $D_i(x,y)$, as shown in Fig.~\ref{fig:water_animate}. To guarantee the conservation of the water volume, we renormalize each filter such that
\begin{equation}
\iint_V D_i(x,y) \,dx\,dy=0,
\label{eqn:conserve}
\end{equation}
where $V=[0,1]\times[0,1]\subset\mathbb{R}^2$ for each $i=1,2,...,d$. Thus we now have a tank of water that is constantly being perturbed by input signal $\mathbf{s}_+$ while preserving its total volume.

The simulation of this perturbed shallow water system is done by a modified Lax-Wendroff method \cite{machalinska2014lax}. The method preserves the second order spatial and temporal accuracy even with the presence of the three source terms in the partial differential equations ($p$, $buh$, and $bvh$). The viscous coefficient $b$ is empirically set to $0.3$. In this finite difference method, we discretize the wave surface into a $128\times 128$ grid, and integrate it with time step $\Delta t= 0.03$. 

\begin{table}
    \centering
    \resizebox{0.95\columnwidth}{!}{%
    \begin{tabularx}{0.95\linewidth}{c c c c}
   \Xhline{0.8pt}
Attractors    &   Equations  & LS & LD     \\
   \hline
Sprott N
    &  $\begin{aligned}[0.65\linewidth]
        \dot{x} &= -10y \\
        \dot{y} &= 5x + 5 z^2 \\
        \dot{z} &= 5+5y-10z
        \end{aligned}$    
    &   $\begin{aligned}[0.2\linewidth]
        0.406\\
        0.001\\
        -10.412
        \end{aligned}$        
    &   2.0391 \\
    \Xhline{0.12pt}
R\"ossler
    &  $\begin{aligned}[0.65\linewidth]
        \dot{x} &= 5y-5z \\
        \dot{y} &= 5x + 2.5 y \\
        \dot{z} &= 5xz - 20 z
        \end{aligned}$    
    &   $\begin{aligned}[0.2\linewidth]
        0.612\\
        0.000\\
        -14.524
        \end{aligned}$        
    &   2.0422 \\
    \Xhline{0.12pt}
Halvorsen
    &  $\begin{aligned}[0.65\linewidth]
        \dot{x} &= -1.4x - 4y -4z - y^2 \\
        \dot{y} &= -1.4y -4z -4x -z^2 \\
        \dot{z} &= -1.4z -4x -4y -x^2
        \end{aligned}$    
    &   $\begin{aligned}[0.2\linewidth]
        0.668\\
        0.011\\
        -4.874
        \end{aligned}$        
    &   2.1393 \\
    \Xhline{0.12pt}
Lorenz
    &  $\begin{aligned}[0.65\linewidth]
        \dot{x} &= -10x + 10y \\
        \dot{y} &= 28x -y -xz \\
        \dot{z} &= -8z/3 + xy
        \end{aligned}$    
    &   $\begin{aligned}[0.2\linewidth]
        0.931\\
        -0.017\\
        -14.588
        \end{aligned}$        
    &   2.0627 \\
    \Xhline{0.12pt}
Sprott B
    &  $\begin{aligned}[0.65\linewidth]
        \dot{x} &= 8 yz \\
        \dot{y} &= 8 x -8y \\
        \dot{z} &= 8 - 8 xy
        \end{aligned}$    
    &   $\begin{aligned}[0.2\linewidth]
        1.652\\
        0.000\\
        -9.646
        \end{aligned}$        
    &   2.1713 \\
    \Xhline{0.12pt}
Thomas
    &  $\begin{aligned}[0.65\linewidth]
        \dot{x} &= -1.85x+10\sin(y) \\
        \dot{y} &= -1.85y+10\sin(z) \\
        \dot{z} &= -1.85z+10\sin(x)
        \end{aligned}$    
    &   $\begin{aligned}[0.2\linewidth]
        0.564\\
        -0.036\\
        -6.080
        \end{aligned}$        
    &   2.0869 \\
    \Xhline{0.8pt}
\end{tabularx}
    }
    \caption{\textbf{Mathematical form of the chaotic attractors studied in this work.} Here we provide the equations for the $6$ chaotic attractors, together with the numerically calculated Lyapunov spectrum (LS) as well as the Lyapunov dimension (LD) of the attractor calculated using the Yorke-Kaplan conjecture.}
    \label{tab:attractors}
    \end{table}

Starting from the initial quiescent wave surface $h(x,y,0)=1$, we drive this dissipative water wave system by the $d$-dimensional mixed trajectory $\mathbf{s}_+(t)$. After a transient period ($T_{\text{dump}}=600$) that is long enough to wash out the effect of the initial condition, we record the water's reaction to the mixed signal. To reduce the amount of data recorded, we sparsely measure the deviation of water elevation from the equilibrium height $h=1$ at $2000$ randomly selected locations, denoted as $\mathbf{h}(t)\in\mathbb{R}^{2000}$. Then, with the available separated trajectories during the training period, we construct output functions that map $\mathbf{h}(t)$ into separated signals $[\mathbf{s}_a(t),\mathbf{s}_b(t)]$. Although many other forms of output functions may also work, we adopt the following nonlinear form with a $\tanh$-type saturation on the quadratic and cubic terms,
\begin{equation}
\begin{bmatrix}
   \mathbf{s}_a(t)\\
   \mathbf{s}_b(t) 
   \end{bmatrix} = \mathbf{W}\begin{bmatrix}
   \mathbf{h}(t)-\mathbf{1} \\
   \tanh((\mathbf{h}(t)-\mathbf{1})^2) \\
   \tanh((\mathbf{h}(t)-\mathbf{1})^3) 
   \end{bmatrix}
\label{eqn:output}
\end{equation}
where $\mathbf{W}\in\mathbb{R}^{6\times6000}$ is the coefficient matrix of the nonlinear output function. With the recorded $\mathbf{h}$ and the available $\mathbf{s}_a$ and $\mathbf{s}_b$ during the training phase ($T_{\text{tran}}=600$ with $20000$ time points), the output weight matrix $\mathbf{W}$ is calculated by the least squares method with the Tikhonov regularization, $\alpha=0.001$. We note that other output functions that outperform this one should exist. However, the purpose of this simulation is to demonstrate that CSS is indeed solvable by such an intermediate system, rather than to develop an optimal design.

\begin{figure}
	\centering
	\includegraphics[width=0.49\textwidth]{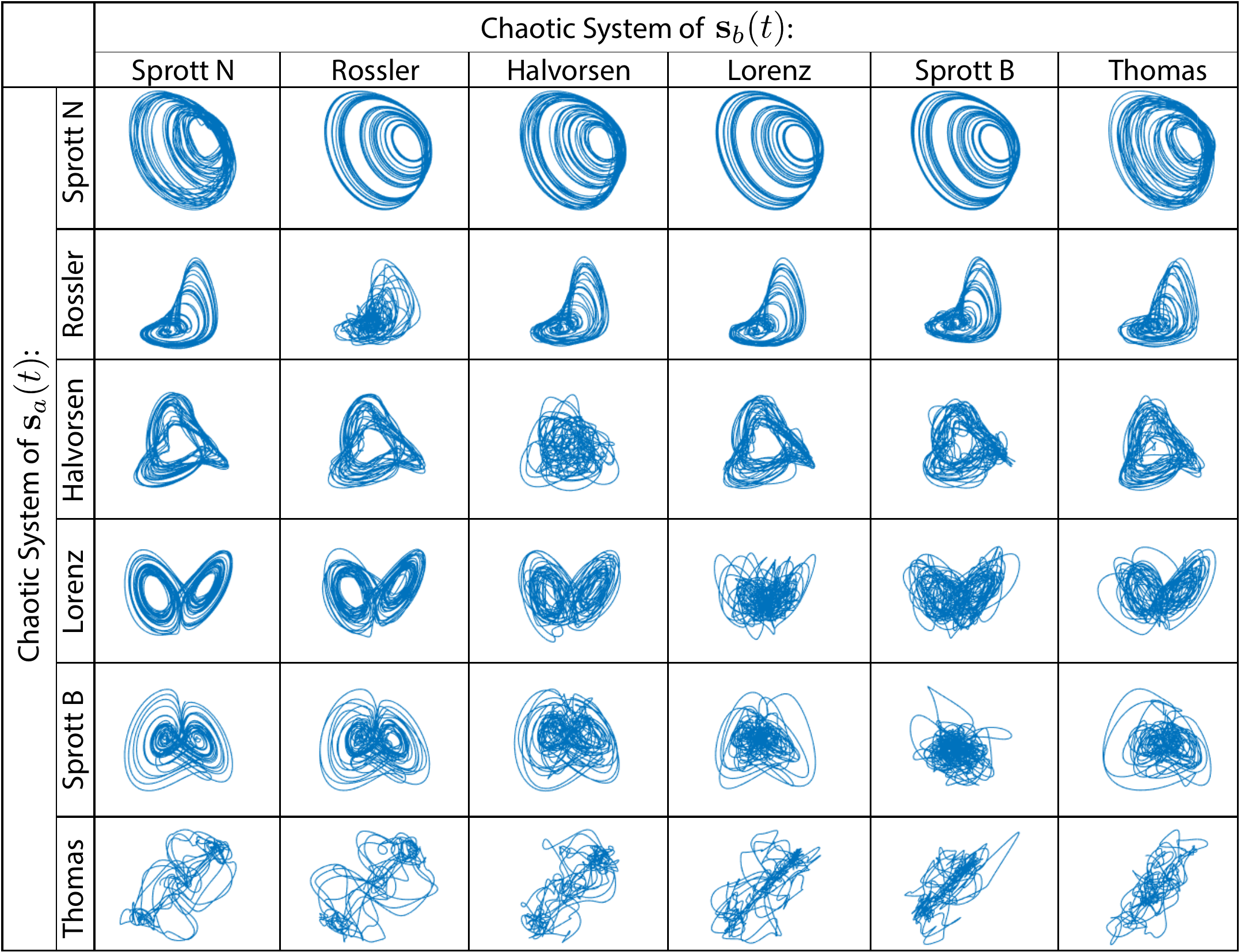}
	\caption{\textbf{Signal separation.} Here we show the signal $\mathbf{s}_a$ separated by a post-training water tank from the mixed signal $\mathbf{s}_+=\mathbf{s}_a+\mathbf{s}_b$, where $\mathbf{s}_a$ and $\mathbf{s}_b$ are the distinct trajectories generated by two of the six chaotic systems listed in Tab.\ref{tab:attractors}, respectively.}
	\label{fig:demixed}
\end{figure}
In Fig.~\ref{fig:water_animate} we show a schematic of the intermediate system separating a mixed signal that is a summation of a Lorenz trajectory and a R\"ossler trajectory. Given the $6$ distinct chaotic systems listed in Tab.~\ref{tab:attractors}, we train and test the separation performance of a shallow water system driven by $6^2-6*(6-1)/2=21$ mixed signals. Each mixed signal is the sum of two chaotic trajectories, $\mathbf{s}_a(t)$ and $\mathbf{s}_b(t)$, that are from the $i$th and the $j$th chaotic system, respectively, for $1\leq i\leq j\leq 6$. To test the system's performance in separating each mixed signal, we reinitialize the water at quiescence, and drive it with a new mixed signal. The separated signal $\mathbf{s}_a$ from the post-training water system, after a transient period ($T=600$), are plotted in Fig.~\ref{fig:demixed}. Specifically, the trajectory on row $i$ and column $j$ is the separated $\mathbf{s}_a$, where the mixed signal is a summation $\mathbf{s}+=\mathbf{s}_a+\mathbf{s}_b$ with $\mathbf{s}_a$ and $\mathbf{s}_b$ from system $i$ and system $j$. For cases where $i=j$, we ensure that the two trajectories being mixed are distinct, i.e., $\mathbf{s}_a(t)\neq\mathbf{s}_b(t)$, by initializing their initial conditions differently.  

\begin{figure}
	\centering
	\includegraphics[width=0.4\textwidth]{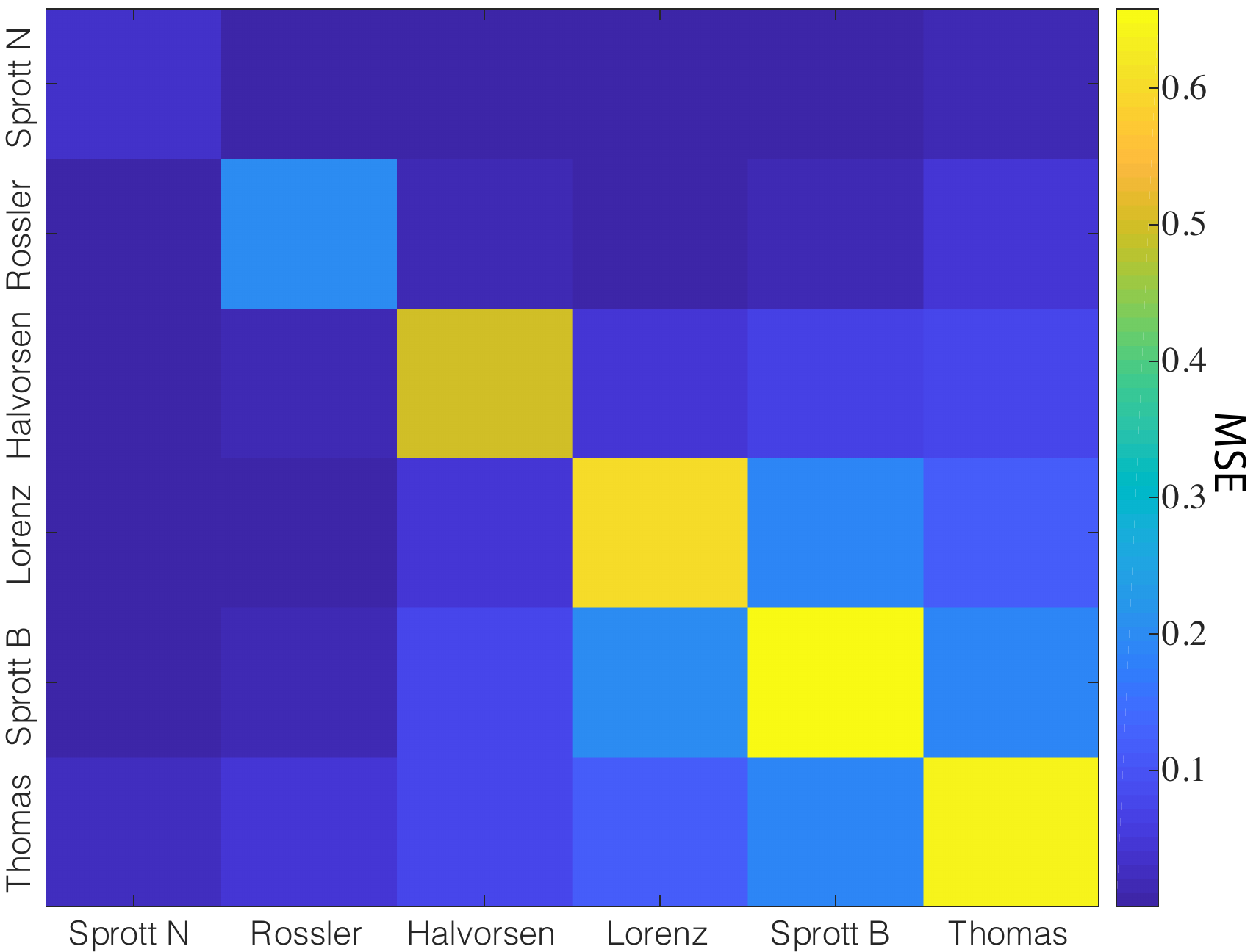}
	\caption{The mean squared errors (MSEs) of the water-tank separated $\mathbf{s}_a$ signals as shown in Fig.~\ref{fig:demixed}. Each $(i,j)$-th element represents the MSE of reconstruction of the signal from system $i$ from a mixed signal of attractors $i$ and $j$.}
	\label{fig:mse}
\end{figure}

By visually collating the separated signals in Fig.~\ref{fig:demixed} and the $6$ chaotic attractors in Fig.~\ref{fig:attractor}, we note that the performance of the separation varies across chaotic systems. Specifically, the Sprott N trajectories and the R\"ossler trajectories separated from a mixture with other systems seem to have much higher quality compared with others (see the first and the second rows in Fig.~\ref{fig:demixed}). To quantify performance, we calculate the mean squared error (MSE) between the actual trajectory and the separated one during the post-training period following $T=600$ (see Fig.~\ref{fig:mse}). We do not find a concrete relationship between the separation performances and the Lyapunov dimensions of the chaotic attractors. However, we note that the quality of the separation appears particularly poor when the two source signals are trajectories from the same chaotic system (see the diagonal line in Fig. ~\ref{fig:demixed}). In the next section, we explain the underlying mechanism behind this supervised CSS, and give an explanation for the diminished performance when signals are taken from the same attractor.

\section{\label{sec:Mechanism}Underlying mechanism of supervised chaotic source separation}
We notice that the chaotic source separation (CSS) problem is essentially a nonlinear state-observer problem, and the intermediate system plays the role of the state-observer. To explicitly state this role, we rewrite the dynamical equations of the two source systems by combining them into a single $2d$-dimensional autonomous dynamical system, denoted by
\begin{equation}
\dot{\mathbf{x}} = 
  \mathbf{f}_{ab}(\mathbf{x})
  =  
  \begin{bmatrix}
    \mathbf{f}_a(\mathbf{s}_a)\\
    \mathbf{f}_{b}(\mathbf{s}_b)
  \end{bmatrix},
\label{eqn:combined_system}
\end{equation}
where 
\begin{equation}
\mathbf{x}(t) \equiv  
  \begin{bmatrix}
    \mathbf{s}_{a}(t)\\
    \mathbf{s}_{b}(t)
  \end{bmatrix},
\label{eqn:combined_state}
\end{equation}
is the direct sum of the states of the two source systems. Thus, the simultaneous mixed signal $\mathbf{s}_+(t)$, a summation of the two source systems, can be considered as an output from the combined system depicted in Eq.~(\ref{eqn:combined_system}), i.e.,
\begin{equation}
\mathbf{y} = \mathbf{g}(\mathbf{x}),
\label{eqn:mixed_out}
\end{equation}
where $\mathbf{g}(\mathbf{s}_a\times\mathbf{s}_b) \equiv \mathbf{s}_a+\mathbf{s}_b = \mathbf{s}_+ \in\mathbb{R}^d$. The CSS problem can then be considered as the problem of uncovering $\mathbf{x}$ in Eq.~(\ref{eqn:combined_system}) through $\mathbf{y}$ in Eq.~(\ref{eqn:mixed_out}), which is the measurement of the combined system.

More precisely, a state-observer is a system that estimates the full state variable of the combined system, $\mathbf{x}$, by considering the measured $\mathbf{y}$ from the combined system. In our method, the state-observer is the intermediate system driven by $\mathbf{y}$, i.e., the perturbed water tank. Such a dynamical state-observer solves the CSS problem by capitalizing on the power of invertible generalized synchronization \cite{lu2018parsimonious,lu2018attractor,lu2017reservoir}.

Specifically, as the intermediate system is driven by the measured $\mathbf{y}$, it evolves nonautonomously according to
\begin{equation}
\frac{d}{dt}\mathbf{r} = \mathbf{f}_r(\mathbf{r},\mathbf{y}).
\label{eqn:observer}
\end{equation}
After a transient period, if the state-observer and the combined system exhibit invertible generalized synchronization, the state of the intermediate system is then uniquely determined by the concurrent state of the combined system $\mathbf{x}$ through an invertible map, i.e., $\mathbf{r}(t)=\phi(\mathbf{x}(t))$. Thus, to estimate the full state of the combined system, one simply needs to train a readout function $\phi^{-1}(\cdot)$ that approximates the inverse of the generalized synchronization function based on the state of the intermediate system $\mathbf{r}$. Notice that invertible generalized synchronization is a property that emerges from particular choices of both the intermediate system and the combined system. As such, we do not find a general principle of designing an intermediate system that guarantees invertible generalized synchronization with any combined chaotic system.

After elucidating the connection between the CSS problem and the state-observer problem, we emphasize that it is only when the full state is observable that such an invertible generalized synchronization function $\phi(\cdot)$ can exist. In other words, the combined system (Eq.~(\ref{eqn:combined_system})) has to be observable through the output function (Eq.~(\ref{eqn:mixed_out})). The classical work of Kalman has discussed the observability of linear dynamical systems (see \cite{ogata2002modern}). In our case, however, the combined dynamical system is nonlinear and autonomous. The necessary and sufficient condition for such a combined system (Eq.~(\ref{eqn:combined_system})) to be observable through the measured output $\mathbf{y}$ is discussed by Inouye in \cite{inouye1977observability}.

Specifically, the system is observable if and only if the observability mapping
\begin{equation}
\mathbf{G}_k(\mathbf{x}) = 
  \begin{bmatrix}
    \mathbf{g}_0(\mathbf{x})\\
    ...\\
    \mathbf{g}_{k-1}(\mathbf{x})
  \end{bmatrix}
\label{eqn:observability_matrix}
\end{equation}
is univalent \cite{inouye1977observability}, where the entries are defined as
\begin{subequations}
\begin{align}
\mathbf{y}(t) & = \mathbf{g}(\mathbf{x}(t)) =  \mathbf{g}_0(\mathbf{x}), \\
\frac{d}{dt}\mathbf{y}(t) & = \frac{\partial \mathbf{g}_0}{\partial \mathbf{x}} \mathbf{f}_{ab}(\mathbf{x}(t)) = \mathbf{g}_1(\mathbf{x}), \\
\frac{d^2}{dt^2}\mathbf{y}(t) & = \frac{\partial \mathbf{g}_1}{\partial \mathbf{x}} \mathbf{f}_{ab}(\mathbf{x}(t)) = \mathbf{g}_2(\mathbf{x}), \\
...\nonumber \\
\frac{d^k}{dt^k}\mathbf{y}(t) & = \frac{\partial \mathbf{g}_{k-1}}{\partial \mathbf{x}} \mathbf{f}_{ab}(\mathbf{x}(t)) = \mathbf{g}_k(\mathbf{x}).
\end{align}
\label{eqn:observability_entry}%
\end{subequations}
When both $\mathbf{f}_{ab}(\cdot)$ and $\mathbf{g}(\cdot)$ are analytic functions on $\mathbb{R}^d$, the system (Eq.~(\ref{eqn:combined_system},\ref{eqn:mixed_out})) is observable if and only if the equations $\mathbf{G}_k(\mathbf{x})=\mathbf{G}_k(\mathbf{x}')$ with $k=1,2,...$ imply only the trivial solution $\mathbf{x}=\mathbf{x}'$.

With the necessary and sufficient condition for observability, we can now investigate whether the CSS problem can be solved providing the measured $\mathbf{s}_+$ when the two source chaotic systems share the same dynamical equation, such that $\mathbf{f}_a(\cdot)=\mathbf{f}_b(\cdot)=\mathbf{f}(\cdot)$. To answer this question, we define
\begin{subequations}
\begin{align}
\mathbf{x} &=\begin{bmatrix}\mathbf{s}\\ \mathbf{s}'\end{bmatrix},\\
\mathbf{x}' &=\begin{bmatrix}\mathbf{s}'\\ \mathbf{s}\end{bmatrix},
\end{align}
\label{eqn:pair_of_x}%
\end{subequations}
where $\mathbf{s}\neq\mathbf{s}'$are distinct trajectories generated by $\dot{\mathbf{s}}=\mathbf{f}(\mathbf{s})$. We then rewrite Eq.~(\ref{eqn:combined_system}) as 
\begin{equation}
\frac{d}{dt}
\begin{bmatrix}
\mathbf{s}\\
\mathbf{s}'
\end{bmatrix}=
\mathbf{f}_{ab}\left(
\begin{bmatrix}
\mathbf{s}\\
\mathbf{s}'
\end{bmatrix}\right)
=
\begin{bmatrix}
\mathbf{f}(\mathbf{s})\\
\mathbf{f}(\mathbf{s}')
\end{bmatrix},
\label{eqn:same_combine}
\end{equation}
and rewrite Eq.~(\ref{eqn:mixed_out}) as
\begin{equation}
\mathbf{s}_+ = \mathbf{g}\left(
\begin{bmatrix}
\mathbf{s}\\
\mathbf{s}'
\end{bmatrix}\right)
=\mathbf{s}+\mathbf{s}'.
\label{eqn:mixed_out_same}
\end{equation}
By substituting Eqs.~(\ref{eqn:pair_of_x}-\ref{eqn:mixed_out_same}) into Eqs.~(\ref{eqn:observability_matrix}-\ref{eqn:observability_entry}), we observe that for any $k$, $\mathbf{g}_k(\mathbf{x})=\mathbf{g}_k(\mathbf{x}')$ even if $\mathbf{x}\neq\mathbf{x}'$, suggesting that the observability mapping is not univalent. Hence, we explain why the CSS performance along the diagonal in Figs.~\ref{fig:demixed} and \ref{fig:mse} tends to be worse compared to the off-diagonal counterparts in the same row.

\begin{figure*}
	\centering
	\includegraphics[width=0.98\textwidth]{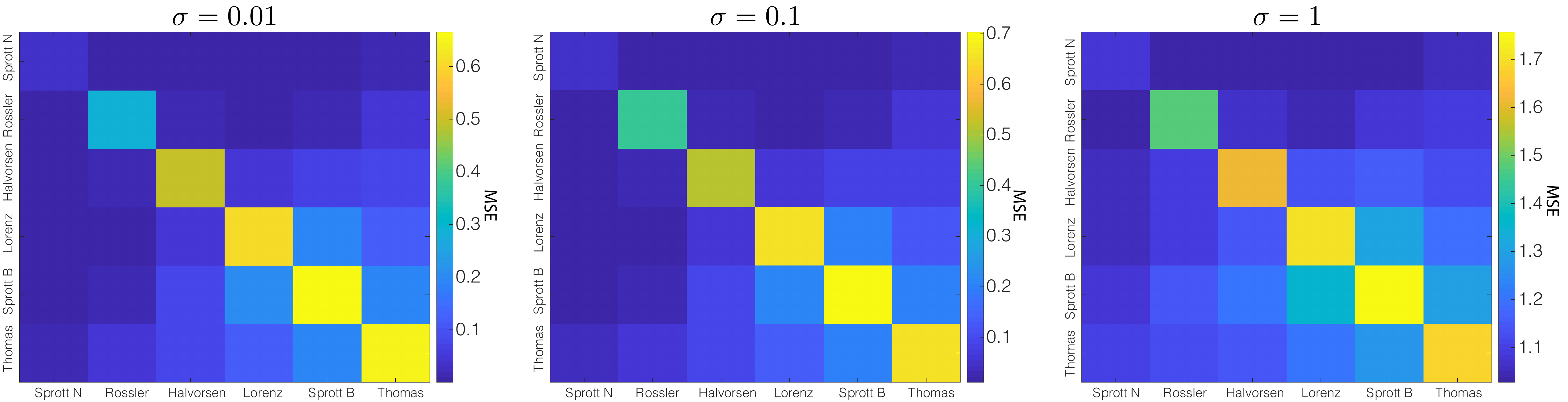}
	\caption{The mean squared errors (MSEs) of the signals $\mathbf{s}_a$ separated by the water-tank when the measurements of the system are corrupted by different levels of noise ($\sigma$): $\sigma=0.01$ (\emph{left}), $\sigma=0.1$ (\emph{center}), and $\sigma=1$ (\emph{right}). Note that the MSE values are similar to those shown in the noise-free simulations displayed in Fig.~\ref{fig:demixed}.}
	\label{fig:mse_noise}
\end{figure*}

\section{\label{sec:Additional}Testing Robustness and Generalizability}
\subsection{Robustness to noise}
To investigate how the CSS performance changes when the source signals are corrupted by observation noise, we modify the simulations in Sec.~\ref{sec:model}. Specifically, we consider that the measured source signals are $\mathbf{s}'_a(t) = \mathbf{s}_a(t)+\sigma\xi^a(t)$ and $\mathbf{s}'_b(t) = \mathbf{s}_b(t)+\sigma\xi^b(t)$, and hence the mixed signal is $\mathbf{s}'_+(t) = \mathbf{s}'_a(t)+\mathbf{s}'_b(t)$, where $\sigma\geq 0$ is the noise strength, and $\xi^{a/b}(t)$ are the white noise terms. By comparing Figs.~\ref{fig:mse_noise} to Fig.~\ref{fig:mse}, we find that the MSE does not significantly increase until the noise strength reaches above $0.1$.

\begin{figure}
	\centering
	\includegraphics[width=0.48\textwidth]{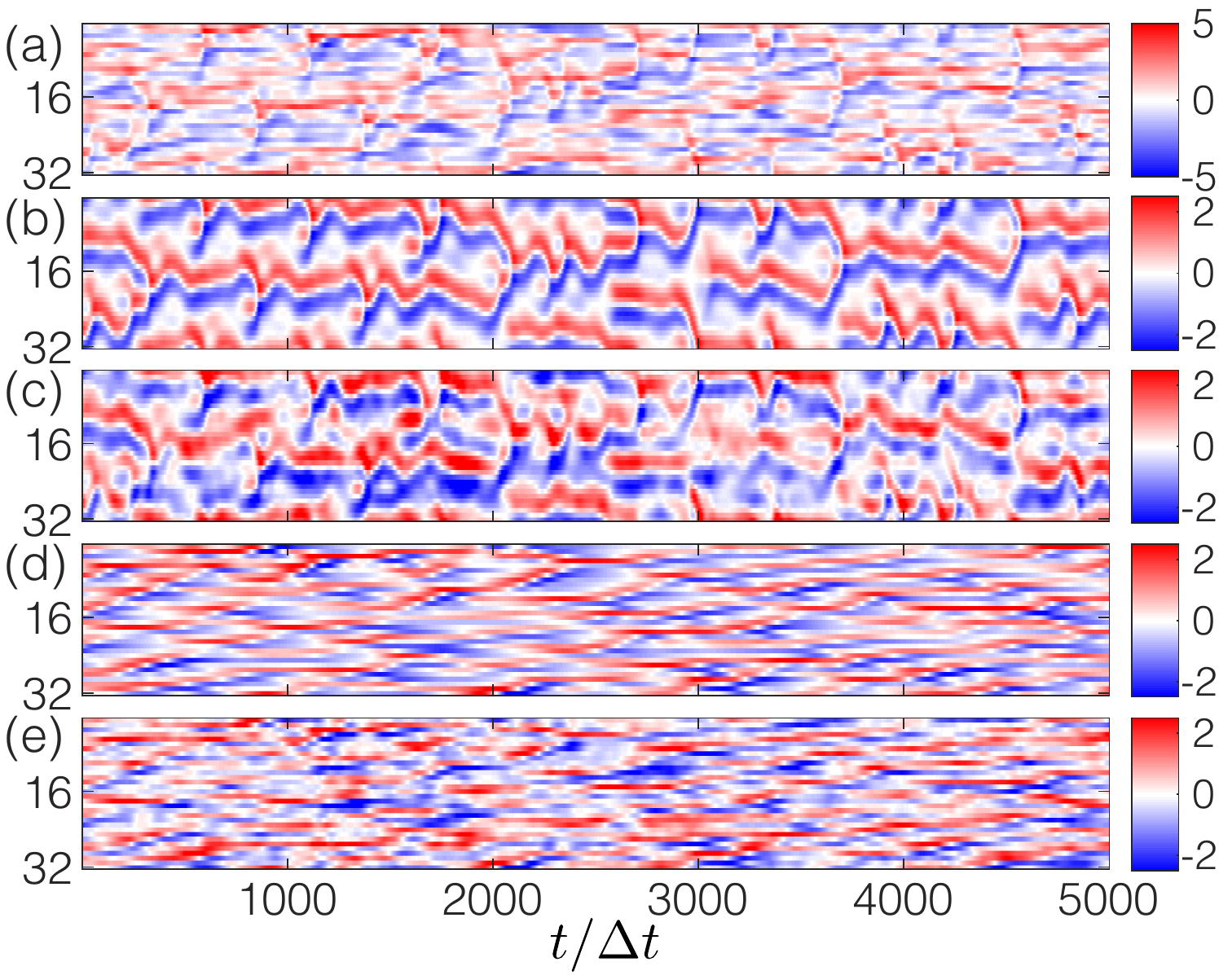}
	\caption{After training, a tank of water can successfully solve the CSS problem with high-dimensional chaotic signals. \emph{(a)} The mixed signal that is the sum of the KS signal \emph{(b)} and the Lorenz 96 signal \emph{(d)}. The estimated KS and Lorenz 96 signals generated by the water tank are shown in panels \emph{(c)} and \emph{(e)} respectively. Signal values are encoded by color.}
	\label{fig:high_Dim}
\end{figure}

\subsection{Generalizability to high dimensional chaotic signals}
Heretofore, we have only tested CSS on source signals that are $3$-dimensional. We now address the question of whether a water tank can be trained to deal with high-dimensional chaotic signals. Accordingly, we employ the Kuramoto-Sivashinsky (KS) system and the Lorenz 96 system as the two chaotic source systems.

We obtain two $32$-dimensional source time series by i) integrating the standard KuramotoâSivashinsky equation~\cite{kuramoto1978diffusion,kassam2005fourth}
\begin{equation}
y_t = -yy_x-y_{xx}-y_{xxxx},
\label{eqn:ks}
\end{equation}
in region $0\leq x<L=22$ (discretized into $32$ evenly spaced grid points) with a periodic boundary condition and time resolution $\Delta t = 1/16$; and ii) integrating the Lorenz 96 equations~\cite{lorenz1996predictability}
\begin{equation}
\frac{dx_i}{dt} = (x_{i+1}-x_{i-2})x_{i-1}-x_i + 8,
\label{eqn:Lorenz96}
\end{equation}
with time resolution $\Delta t = 0.001$ and also a periodic boundary condition where $i=1,2,...,32$. As in the previous simulations, we preprocess the source signals such that each of their variables has mean zero and unit variance along the time axis. The mixed signal (Fig.~\ref{fig:high_Dim} (a)) is then the sum of the two processed source signals (Figs.~\ref{fig:high_Dim} (b,d)).

In Figs.~\ref{fig:high_Dim} (c,e) we show that a tank of water, after being trained, indeed can estimate the high dimensional source signals. For this simulation, we utilize a spatial discretization of $256$-by-$256$, which is finer than the low-dimensional case. The viscous drag coefficient is set to be $b=0.6$. While these parameter choices are sufficient for this demonstration, further parameter optimization could lead to better performance for this or other systems.

\section{\label{sec:disc}Discussion}
Complementing previous studies on source separation problems \cite{karhunen1995nonlinear,stetter2000principal,comon2010handbook,pham1996blind,cichocki2006new,wang2005musical,phon2013handling,phon2013handling,he2018single,sawada2019review,sharma2015blind,han2012classification,mogami2018independent,wang2018supervised,lu2018parsimonious,krishnagopal2019separation}, we show that separation of signals from a mixture of chaotic trajectories can be considered as a nonlinear state observer problem. With this realization, we propose to solve the problem by employing and training an intermediate system that is continuously driven by the mixed signal. We extend earlier studies where CSS is solved by recurrent neural networks \cite{lu2018parsimonious,krishnagopal2019separation}, and we show that even a tank of water under this proposed framework can solve the CSS problem. By making the connection between the CSS problem and the nonlinear state-observer problem, we explain the reason why separating two signals generated from the same chaotic system tends to be difficult.

We note that in this paper, we only consider mixed signals that are sums of two chaotic trajectories. Yet, our method can be applied to other mixing equations, or mixtures of more than two chaotic trajectories. However, we do expect the method to perform less well when the mixture is more complicated or contains more than two source systems. Future studies could seek principles that guarantee the design of a better intermediate system for different chaotic signals.

\section*{Acknowledgements}
We acknowledge support from the John D. and Catherine T. MacArthur Foundation, the Alfred P. Sloan Foundation, the ISI Foundation, the Paul Allen Foundation, the Army Research Laboratory (W911NF-10-2-0022), the Army Research Office (Bassett-W911NF-14-1-0679, Grafton-W911NF-16-1-0474, DCIST-W911NF-17-2-0181), the Office of Naval Research, the National Institute of Mental Health (2-R01-DC-009209-11, R01-MH112847, R01-MH107235, R21-M MH-106799), the National Institute of Child Health and Human Development (1R01HD086888-01), National Institute of Neurological Disorders and Stroke (R01 NS099348), and the National Science Foundation (BCS-1441502, BCS-1430087, NSF PHY-1554488 and BCS-1631550). The content is solely the responsibility of the authors and does not necessarily represent the official views of any of the funding agencies.
%\section*{References}
%\bibliographystyle{}
\bibliography{citation}
\end{document}